\documentclass[11pt, a4paper]{article}

\usepackage{anysize}
\marginsize{3cm}{3cm}{1.5cm}{1.5cm}
\usepackage[T1]{fontenc}
\usepackage{inputenc}

\usepackage{amsfonts}
\usepackage{amssymb}
\usepackage{graphics}
\usepackage{epsfig}
\usepackage{graphicx}
\usepackage{epsfig}
\usepackage{amssymb}
\usepackage{amsfonts}

\usepackage{amsmath}
\usepackage{xcolor}
\usepackage{float}
\usepackage{enumerate}
\usepackage{slashed}
\usepackage{hyperref}
\hypersetup{colorlinks=true,linkcolor=blue,citecolor=blue} 

\numberwithin{equation}{section}

\newcommand {\beq} {\begin{equation}}
\newcommand {\eeq} {\end{equation}}
\newcommand{\bea}{\begin{eqnarray}}
\newcommand{\eea}{\end{eqnarray}}
\newcommand{\bit}{\begin{itemize}}
\newcommand{\eit}{\end{itemize}}
\def\nl{\nonumber \\}

\def\a{\alpha}

\def\b{\beta}

\def\p{\partial}

\def\le{\left(}
\def\ri{\right)}

\def\beq{\begin{equation}}
\def\eeq{\end{equation}}

\begin{document}

\begin{titlepage}

\begin{flushright}

\end{flushright}
\bigskip
\begin{center}
{\LARGE  {\bf
Complexity and action for warped AdS black holes
  \\[2mm] } }
\end{center}
\bigskip
\begin{center}
{\large \bf  Roberto  Auzzi$^{a,b}$},
 {\large \bf Stefano Baiguera$^{c}$},
  {\large \bf Matteo Grassi$^{a}$}, \\
  {\large \bf Giuseppe Nardelli$^{a,d}$}  {\large \bf and } 
    {\large \bf Nicol\`o Zenoni$^{a}$}
\vskip 0.20cm
\end{center}
\vskip 0.20cm 
\begin{center}
$^a${ \it \small  Dipartimento di Matematica e Fisica,  Universit\`a Cattolica
del Sacro Cuore, \\
Via Musei 41, 25121 Brescia, Italy}
\\ \vskip 0.20cm 
$^b${ \it \small{INFN Sezione di Perugia,  Via A. Pascoli, 06123 Perugia, Italy}}
\\ \vskip 0.20cm 
$^c${ \it \small{Universit\`a degli studi di Milano Bicocca and INFN, 
Sezione di Milano - Bicocca, \\ Piazza
della Scienza 3, 20161, Milano, Italy}}
\\ \vskip 0.20cm 
$^d${ \it \small{TIFPA - INFN, c/o Dipartimento di Fisica, Universit\`a di Trento, \\ 38123 Povo (TN), Italy} }
\\ \vskip 0.20cm 
E-mails: roberto.auzzi@unicatt.it, s.baiguera@campus.unimib.it,  \\
matteogrs@gmail.com, giuseppe.nardelli@unicatt.it, \\
zenon94@hotmail.it
\end{center}
\vspace{3mm}

\begin{abstract}
The Complexity=Action conjecture is studied for black holes in Warped AdS$_3$ space,
realized as solutions of Einstein gravity plus matter. The time dependence
of the action of the Wheeler-DeWitt patch is investigated,
 both for the non-rotating and the rotating case.
The asymptotic growth rate is found to be
equal to the Hawking  temperature times the Bekenstein-Hawking entropy; this is 
in agreement with a previous calculation done using the Complexity=Volume conjecture.
\end{abstract}

\end{titlepage}

\section{Introduction}

The AdS/CFT correspondence gives us a non-perturbative
formulation of quantum gravity for a class of spacetimes 
 with negative curvature and AdS asymptotic. Despite many evidences 
 for the validity of the correspondence, it would be  desirable to improve
 our understanding about how the spacetime geometry emerges out of 
 the quantum field theory degrees of freedom living in the boundary.
 Quantum information concepts seem somehow to encode
 non trivial geometric properties of the gravitational theory in the bulk.
 For example, the area of minimal surface in AdS is dual to 
 the entanglement entropy of the boundary subregion
 \cite{Ryu:2006bv,Casini:2011kv,Lewkowycz:2013nqa}.
 However, the precise mechanism by which the dual bulk spacetime 
 geometry emerges out of the boundary quantum field theory
 is still not understood. 
 
 Entropy is a crucial quantity in order to describe classical and quantum aspects 
 \cite{Bekenstein:1973ur,Bardeen:1973gs} of Black  Holes (BHs). However,
 it does not seem the right dual quantity in order to describe 
 the Einstein-Rosen Bridge (ERB) in the interior of a two-sided Kruskal BH.
 In the AdS/CFT correspondence, a two sided eternal BH is dual to a thermofield
doublet state, in which the two conformal field theories living on the left and right boundaries
are entangled  \cite{Maldacena:2001kr}. Taking the two boundary times going in the same direction,
this entangled state is time-dependent \cite{Hartman:2013qma}, and the geometry of the ERB connecting
the two sides grows linearly with time. The ERB continues
 to grow for a much  longer timescale compared to the thermalization time,
 and so entropy does not provide us with a good dual quantity for this process.

 Motivated by the need to find a boundary dual to such behavior,
  recently a new quantum information tool has joined 
 the discussion: computational complexity \cite{Susskind:2014rva,Susskind:2014moa}. 
 For a quantum-mechanical system, it is defined
 as the minimal number of basic unitary operation which are needed in order to prepare
 a given state starting from a simple reference state.
 A proper definition of complexity in quantum field theory 
 has several subtleties, including the choice of the reference state
 and of the allowed set of elementary quantum gates and  the allowed amount of tolerance 
 which is introduced in order to specify the accuracy with which the state should be produced.
 Recently, concrete calculations have been performed in the case of free field theories
  \cite{Jefferson:2017sdb,Chapman:2017rqy,Hashimoto:2017fga,Kim:2017qrq,Khan:2018rzm,Hackl:2018ptj}.
  Another interesting approach to complexity  \cite{Caputa:2017yrh,Bhattacharyya:2018wym}
   in quantum field theory  uses tensor networks \cite{Swingle:2009bg} in connection with the
   Liouville action.  Related papers about general aspects of complexity in 
   field theory include \cite{Yang:2018nda, Hashimoto:2018bmb}.

Two different gravity dual of the quantum complexity of a state have been proposed so far:
the complexity=volume (CV) \cite{Susskind:2014rva,Susskind:2014moa,Stanford:2014jda}
and the complexity=action (CA) \cite{Brown:2015bva,Brown:2015lvg}  conjectures.
 In the CV conjecture, complexity is proportional to the volume $V$ of 
 a maximal codimension one sub-manifold hanging from the boundary.
 In the CA conjecture, complexity $\mathcal{C}$ is proportional to
 the action $I$ evaluated in the causal diamond of a   boundary section
 at constant time, which is called Wheeler-DeWitt (WDW) patch:
 \beq
 \mathcal{C} = \frac{I}{\pi \hbar } \, ,
 \eeq
In this case the action has several contributions
beyond the traditional bulk Einstein-Hilbert  (EH) and boundary Gibbons-Hawking-York (GHY) terms:
in particular surface joint contributions \cite{Hayward:1993my,Lehner:2016vdi}
turn out to be important in order to compute the full time dependence of 
the WDW action \cite{Lehner:2016vdi,Chapman:2016hwi,Carmi:2017jqz,Kim:2017qrq}.
Moreover, ambiguities due to contributions to the action from null
surfaces \cite{Neiman:2012fx,Parattu:2015gga} are also present;
these ambiguities do not affect the late-time limit of complexity,
which can be computed just from the EH and GHY terms in the action
 \cite{Brown:2015bva,Brown:2015lvg}. 

The CA and CV conjectures have been recently investigated in several AdS/CFT
settings: for example for rotating/charged BHs in several dimensions \cite{Cai:2016xho},
for  spacetime singularities \cite{Barbon:2015ria,Bolognesi:2018ion},
for the  soliton \cite{Reynolds:2017jfs}, in the Vaidya spacetime
\cite{Moosa:2017yvt,Moosa:2017yiz,Chapman:2018dem,Chapman:2018lsv}
and in theories with dilatons \cite{Swingle:2017zcd,An:2018xhv}.

Quantum information has been rather extensively studied for asymptotically AdS spacetimes; 
the understanding that we have for other spacetimes, such as the asymptotically flat or the de Sitter,
 is much more limited, because we have so far very little clues about the
 dual field theory, if it exists.  An interesting ultraviolet 
 deformation of AdS/CFT where we have 
 a good amount of information about the structure of the field theory dual is the
Warped AdS$_3$/CFT$_2$ correspondence
\cite{Anninos:2008fx,Detournay:2012pc,Hofman:2014loa,Jensen:2017tnb}.
This is a duality between gravitational theories
in $2+1$ dimensions in a space with Warped AdS$_3$ asymptotic
and a conjectured class of non-relativistic theories in $1+1$ dimensions,
called Warped Conformal Field Theories (WCFTs), whose symmetry content includes
a copy of the Virasoro and of the $U(1)$ Kac-Moody current algebras.
Recently, several progresses have been made in order to put
this duality on firmer grounds;  for example,
  an analog of Cardy formula was derived in \cite{Detournay:2012pc}.
The issue of entanglement entropy was studied by several authors,
e.g.  \cite{Anninos:2013nja, Castro:2015csg, Azeyanagi:2018har,Song:2016pwx,Song:2016gtd}.  
The CV conjecture was recently  studied in \cite{Auzzi:2018zdu};
in this paper we will instead address the CA conjecture.

The paper is organized as follows: in section \ref{sect-BH-generalities} we review general properties of
 BHs in WAdS space, realized as a solution of
 Einstein gravity plus matter,  and we discuss the null coordinates needed to define the WDW patch.
  In section \ref{sect-action} we consider
 the various contributions to the action, following the approach of  \cite{Lehner:2016vdi}.
 In section \ref{CA} we compute the action
for both the non-rotating and rotating case. We conclude in section \ref{sect-conclu}. Technical details
about the matching with the metric of \cite{Banados:2005da}
 are discussed in appendix \ref{comparison}.
An alternative calculation
 using the approach of \cite{Brown:2015lvg}
 is presented in appendix  \ref{another-way}:
 this is valid just in the late-time limit and agrees
 with the more general calculation presented in section \ref{CA}.


\section{Warped Black Holes in Einstein gravity}
\label{sect-BH-generalities}

We consider the following class of BHs  with Warped AdS$_3 $ asymptotic
 \cite{Moussa:2003fc,Bouchareb:2007yx,Anninos:2008fx}:
\beq
\label{BHole}
\frac{ds^2}{l^2} = dt^2 +\frac{ dr^2}{(\nu^2+3) (r-r_+)(r-r_-)}
+\le 2 \nu r -\sqrt{r_+ r_- (\nu^2+3)}  \ri dt d \theta +  \frac{r}{4}  \Psi d \theta^2 \, ,
\eeq
\beq
 \Psi(r)= 3 (\nu^2-1)  r +(\nu^2+3) (r_+ + r_-) - 4 \nu \sqrt{ r_+ r_- (\nu^2+3)} \, .
\eeq
We introduce $\tilde{r}_0$ as 
\beq
\tilde{r}_0= \max(0, \rho_0) \, , \qquad \rho_0=\frac{4 \nu \sqrt{r_+ r_- (\nu^2+3) }- (\nu^2+3)(r_++r_-)}{3 (\nu^2-1)} \, ,
\eeq
where $\Psi(\rho_0)=0$ and we take the range of variables as follows:
 $ \tilde{r}_0 \leq r  < \infty $, $-\infty < t < \infty$, 
 $ \theta \sim \theta + 2 \pi$ and the horizons are located at $r=r_+, r_-$
with $r_+ \geq r_-$.
These metrics can be obtained by discrete quotients of WAdS$_3$ \cite{Anninos:2008fx};
we take $\nu \geq 1$ in order to avoid closed time-like curves. 
For $\nu=1$ the metric (\ref{BHole})
reduces to the  the  Banados-Teitelboim-Zanelli (BTZ) black hole 
\cite{Banados:1992wn,Banados:1992gq}.
The warping  parameter $\nu$ is related in the holographic dictionary
to the left and right central charges of the boundary WCFT, 
which for Einstein gravity are \cite{Anninos:2008qb}:
\beq
c_L=c_R=\frac{12 l \nu^2 }{G (\nu^2+3)^{3/2}} \, .
\eeq
Temperature and angular velocity of horizon are \cite{Anninos:2008fx}:
\beq
T= \frac{\nu^2+3}{4 \pi l } \,  \frac{r_+ -r_-}{2 \nu r_+ -\sqrt{(\nu^2+3) r_+ r_-} } \, ,
\qquad
\Omega=\frac{2}{(2 \nu r_+ -\sqrt{(\nu^2+3) r_+ r_-}) l } \, .
\eeq

The metric (\ref{BHole}) can be obtained as a
 vacuum solution of Topologically Massive Gravity (TMG)
 \cite{Moussa:2003fc,Bouchareb:2007yx},
New Massive Gravity (NMG) \cite{Clement:2009gq},
general linear combinations of the two mass terms \cite{Tonni:2010gb} and also
in string theory constructions \cite{Compere:2008cw,Detournay:2012dz,Karndumri:2013dca}.
We will be interested to WAdS$_3$ BHs realized as solution of Einstein gravity
with matter. Unfortunately, all the known realizations of WAdS$_3$ BHs 
in Einstein gravity have some pathology in the matter content:
for example, they can be realized as solutions
with perfect fluid stress tensor with spacelike quadrivelocity \cite{Gurses:1994bjn}.

We will use for concreteness the model studied in
 \cite{Banados:2005da,Barnich:2005kq}, which is 
 Chern-Simons-Maxwell electrodynamics 
coupled to Einstein gravity.
In order to have solutions without closed time-like curves,
a wrong sign for the kinetic Maxwell term is needed.
Solutions with positive Maxwell kinetic energy have $\nu^2<1$ and correspond
to G\"odel spacetimes.
We will see that the CA conjecture is so solid that can survive
to unphysical action with ghosts.

In the Einstein gravity case the entropy is given by the area of the horizon:
\beq
S=\frac{l \pi}{4 G} (2 \nu r_+ - \sqrt{r_+ r_- (\nu^2+3)}) \, .
\eeq
and the conserved charges (mass and angular momentum) are \cite{Banados:2005da,Barnich:2005kq,Auzzi:2018zdu}:
\beq
M=\frac{1}{16 G} (\nu^2+3)
 \le  \le r_{-} + r_{+} \ri - \frac{\sqrt{r_{+}r_{-} (\nu^2 +3)}}{\nu} \ri \, ,
 \label{MM}
\eeq
\beq
J=\frac{l}{32 G} (\nu^2+3) 
\le \frac{r_- r_+ (3+5 \nu^2)}{2 \nu}
 -(r_+ + r_-) \sqrt{(3+\nu^2) r_+ r_-} 
 \ri \, .
\label{JJ}
\eeq

\subsection{Null coordinates}

The expression of the metric (\ref{BHole}) in Arnowitt-Deser-Misner (ADM) form is: 
 \beq
 \label{BHole-ADM}
 ds^2=-N^2 dt^2  +\frac{l^4 dr^2}{4 R^2 N^2}
 +l^2 R^2 (d \theta + N^\theta dt)^2
 \, , 
 \eeq
 where
 \beq
 R^2=\frac{r}{4} \Psi \, , \qquad
 N^2=\frac{l^2 (\nu^2+3) (r-r_+)(r-r_-)}{4 R^2} \, ,
 \qquad
 N^\theta= \frac{2 \nu r -\sqrt{r_+ r_- (\nu^2+3)} }{2 R^2} \, .
 \eeq
 It is useful to use a set of null coordinates which delimit the 
 WDW patch. These coordinates were  introduced in  \cite{Jugeau:2010nq}.
We consider a set of null geodesics
which satisfy $(d \theta + N^\theta dt)=0$; then 
 a positive-definite term in the metric (\ref{BHole-ADM})
 saturates to zero, and the null geodesics are given by the constant $u$ and $v$ trajectories:
 \beq
du = dt - \frac{l^2}{2 R N^2} dr \, , \qquad
dv = dt + \frac{l^2}{2 R N^2} dr  \,  .
 \eeq
 The normal one-forms to the WDW null surfaces are given by
$du$ and $dv$; we introduce two vectors $v_\a$, $u_\a$
such that
\beq
dv = v_\a dx^\a \, , \qquad du=u_\a dx^\a \, ,
\label{null-normals}
\eeq
which are normal and tangent to the null surfaces which
delimit the WDW patch.
The corresponding  Eddington-Finkelstein coordinates then are:
\beq
u= t - r^* (r) \, , \qquad
v= t + r^* (r) \, ,  
\label{Kruskalcoordinates}
\eeq
where 
\beq
\frac{dr^*}{dr} = \frac{l^2}{2 R N^2} =\frac{\sqrt{r \, \Psi(r) }}{(\nu^2 +3)(r-r_-)(r-r_+)} \, .
\label{derivatarstar}
\eeq 
The non-rotating case is defined by the condition $J=0$, 
and corresponds  to the following values:
\beq
r_-=0 \, , \qquad  \frac{r_+}{r_-}=\frac{4 \nu^2}{\nu^2+3} \, .
\label{non-rota}
\eeq
In this case the Penrose diagram is the same as the ones for the Schwarzchild BH
in four dimension  \cite{Jugeau:2010nq}. 
In the rotating case, for generic $(r_+,r_-)$, 
the Penrose diagram is  the same as the one of the
Reissner-Nordstr\"om BH.


\subsection{An explicit model}
 
In this section we consider an explicit Einstein gravity model which admits 
 the metric eq.~(\ref{BHole}) as a solution \cite{Banados:2005da}.
The matter content is a gauge field with Chern-Simons and Maxwell terms,
and the bulk part of the action is:
\beq
I_{\mathcal{V}}=  \frac{1}{16 \pi G}  \int d^3 x \left\{\sqrt{g} \left[ 
\left( R + \frac{2}{L^2}\right) -\frac{\kappa}{4} F^{\mu \nu} F_{\mu \nu}
\right] -\frac{\a}{2} \epsilon^{\mu \nu \rho} A_\mu F_{\nu \rho}\right\}= \int d^3 x \sqrt{g} \mathcal{S}
 \, ,
\label{S-bulk}
\eeq
where $\epsilon^{\mu \nu \rho} $ is the Levi-Civita tensorial density.
Here we put a coefficient $\kappa=\pm 1$ in front of  the Maxwell kinetic term.

The equations of motion for the gauge field are
\beq
D_\mu F^{\a \mu} = -\frac{\a}{\kappa}  \frac{ \epsilon^{\a \nu \rho} }{\sqrt{g}} F_{\nu \rho} \, ,
\eeq
while the Einstein equations are
\beq
G_{\mu \nu} - \frac{1}{L^2} g_{\mu \nu} = \frac{\kappa}{2}  T_{\mu \nu} \, , \qquad
T_{\mu \nu} =   F_{\mu \a} F_{\nu}^{\,\,\, \a}-\frac14 g_{\mu \nu}
F^{\a \b} F_{\a \b} \, .
\eeq
We consider the set of coordinates $ (r,t, \theta) $ where the metric assumes the form (\ref{BHole}), and we choose a 
 gauge motivated by the ansatz from \cite{Banados:2005da}:
\beq
A= a dt+ (b+c r) d \theta \, , \qquad F = c \, dr \wedge d \theta \, ,
\label{ansatz per campo di gauge}
\eeq
where $ \lbrace a,b,c \rbrace $ is a set of constants. Thus, the Maxwell equations give:
\beq
\a = \kappa \frac{\nu}{l} \, .
\eeq
From the Einstein equations, we get, independently from $(r_+,r_-)$:
\beq
L = l \sqrt{\frac{2}{3-\nu^2} } \, , \qquad
c =\pm  l \sqrt{\frac{3}{2 } \frac{1-\nu^2}{\kappa}}
\, . 
\label{gaugau}
\eeq
There is conflict between absence of closed time-like  curves
and presence of ghosts ($\kappa=-1$).

Note that the  parameters $a,b$ are not constrained by the equations of motion;
the action itself does not depend on the parameter $b$, but it depends 
explicitly on the gauge parameter $a$ through the Chern-Simons term.
This parameter is important in order to properly define the conserved charge
which gives the mass $M$ \cite{Barnich:2005kq}. Only for a particular value
of $a$ the mass is indeed associated to the
Killing vector $\p/\p t$ and is independent from the $U(1)$
gauge transformations. 
This corresponds to the $\zeta=0$ gauge in  \cite{Banados:2005da}; in our notation it correponds to:
\beq
A_t=a=
\frac{l}{\nu} \sqrt{\frac32} \sqrt{\nu^2-1}  \, .
\label{a-action}
\eeq
The comparison with the solution of  \cite{Banados:2005da} is
discussed in appendix \ref{comparison}.


\section{Evaluating the action}
\label{sect-action}

The action in the WDW patch has several contributions:
\beq
I= I_{\mathcal{V}} + I_{\mathcal{B}} + I_{\mathcal{J}} \, ,
\eeq	
where $I_{\mathcal{V}}$ is the bulk contribution
(see eq. (\ref{S-bulk})),  $I_{\mathcal{B}}$ the
the boundary  term and 
$I_{\mathcal{J}} $ the joint term studied in detail in \cite{Lehner:2016vdi}.

The bulk action integrand $\sqrt{g} \mathcal{S}$ in 
eq. (\ref{S-bulk}) evaluated on the background (\ref{BHole}) and (\ref{ansatz per campo di gauge})
 is constant and independent from the parameters $(r_+,r_-)$:
\beq
I_{\mathcal{V}} =\int dr dt d \theta \frac{\mathcal{I} }{16 \pi G} \, , \qquad
\mathcal{I} = 
- \frac{l}{2} (\nu^2 +3) + \frac{\kappa c^2}{l} - \alpha a c \, .
\eeq

The boundary terms can be written as:
\beq
I_{\mathcal{B}} = I_{\rm GHY} + I_{\mathcal{N}} \, ,
\eeq
where $I_{\rm GHY}$ is the contribution for spacelike and timelike boundaries
(Gibbons-Hawking-York (GHY) term)
and $ I_{\mathcal{N}} $ is the contribution for null boundaries.
The GHY term is:
\beq
I_{\rm GHY} = \frac{  \varepsilon }{8 \pi G}  \int_{\mathcal{B}} d^2 x \, \sqrt{|h|} \, K \, ,
\eeq
where $ \mathcal{B} $ is the appropriate boundary, \emph{h} the induced metric, \emph{K} the extrinsic curvature
and $ \varepsilon$ is equal to $+1$ if the boundary is timelike and $-1$ if it is spacelike.
For  null surface boundaries the contribution to the action is \cite{Neiman:2012fx, Parattu:2015gga, Lehner:2016vdi}
\beq
I_{\mathcal{N}}= \frac{  1 }{8 \pi G}  \int_{\mathcal{B}} \tilde{\kappa} d \lambda d S \, ,
\eeq
where  $\lambda$ parameterizes the null direction of the surface,  $d S$
is the area element of the spatial cross-section orthogonal to the null direction
and $\tilde{\kappa}$ measures the failure of $\lambda$ to be an affine parameter:
if we denote by $k^\a$ the null generator, $\tilde{\kappa}$  is defined by the relation:
$k^\mu D_\mu k^\a = \tilde{\kappa} k^\a$. 
 It turns out that the contribution to the action
$I_{\mathcal{N}}$ is not parameterization-invariant \cite{Parattu:2015gga,Lehner:2016vdi}
 and it can be set to zero using an
affine parameterization for the null direction of the boundary
\cite{Lehner:2016vdi}. 

In the case of joints between spacelike and timelike surfaces, this contribution
was studied in \cite{Hayward:1993my}. The analysis for joints 
between null and timelike, spacelike or another null surface were recently studied in
\cite{Lehner:2016vdi}. 
In the CA calculations done in the next sections, we will use these null joints
contributions several times:
\beq
I_{\mathcal{J}} = \frac{1}{8 \pi G} \int_{\Sigma} d \theta \sqrt{\sigma} \, \mathfrak{a} \, ,
\label{jjoints}
\eeq
where $ \sigma_{ab} $ is the induced metric over the joint (in this case, it is 1-dimensional)
and $ \mathfrak{a} $  depends on the kind of joint.
Let us denote $k^\a$ the future directed null normal to a null surface
(which is also tangent to the surface), $n_\a$ the normal to a spacelike surface 
and $s_\a$ the normal to a timelike surface, both directed outwards the volume of interest. 
In the case of intersection of two null surfaces with normals  $k^\a_1$ and $k^\a_2$:
\beq
\mathfrak{a} = \eta  \log \left| \frac{k_1 \cdot k_2}{2} \right| \, ,
\label{jnn}
\eeq
while in the case of intersection of a null surface with normal $k^\a$
and a spacelike surface with normal $n^\a$ (or a timelike surface with normal $s^\a$):
\beq
\mathfrak{a} = \eta  \log \left|  k \cdot n \right| \, , \qquad
\mathfrak{a} = \eta \log \left|  k \cdot s \right| \, .
\label{jns}
\eeq
In eqs. (\ref{jnn}-\ref{jns}), if  the outward direction
to the region of interest is pointing along the future,
we should set $ \eta= 1 $ 
if the joint lies in the future of the spacetime volume of interest, 
and $ \eta=-1 $ if the joint lies in the past.
If instead the outward direction is pointing along the past,
we should set $ \eta= 1 $ 
if the joint lies in past of the spacetime volume of interest, 
and $ \eta=-1 $ if the joint lies in the future.

Note that  eqs.  (\ref{jnn}) and (\ref{jns})  are slightly ambiguous because
the normalization of a null normal $k^\a$ is ambiguous.
This ambiguity is related to the one due to the null surfaces
and does not affect the late-time limit of the complexity, but just
the finite-time behavior\footnote{These ambiguities could be related to various ambiguities 
of the dual circuit complexity of the quantum state,
such as the choice of the reference state,
the specific set of elementary gates and the amount of tolerance 
that one introduces to describe the accuracy with which the final state should
be constructed. }.
As discussed in \cite{Carmi:2017jqz},
we will partially fix this ambiguity by requiring that the null vector $k^\mu$
have constant scalar product with the boundary time killing vector $\p/\p t$.

\section{Complexity=Action}
\label{CA}

\begin{figure}[h]
\begin{center}
\leavevmode
\epsfig{file=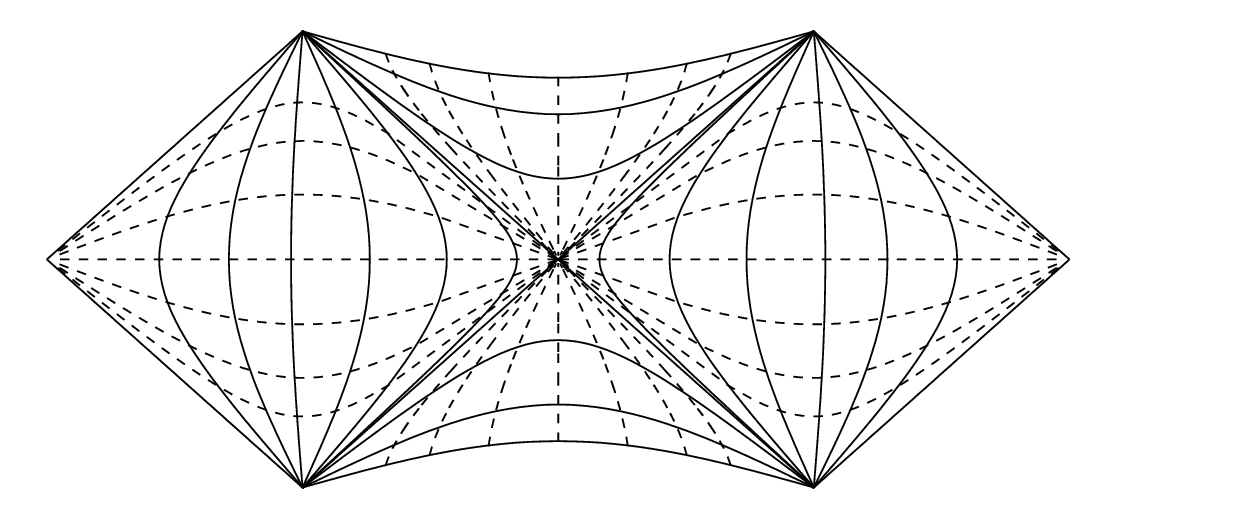, width =5.0in}
\end{center}
\caption{Constant $r$ lines (solid) and constant $t$ lines (dashed) 
of the Penrose diagram in the non-rotating case.}
\label{fig-t}
\end{figure}

The Penrose diagram for the non-rotating case is shown in figure \ref{fig-t},
with some lines at constant $r$ and $t$. 
Both in the rotating and non-rotating cases,
for $r \rightarrow \infty$,  the asymptotic behavior of $r^*(r)$ is
\beq
r^* (r) \approx \frac{ \sqrt{3(\nu^2-1)}}{\nu^2+3} \log r \equiv C  \log r  \, .
\eeq
So we should first fix a cutoff surface at $r= \Lambda$ to make our calculations finite.
The WDW surface is bounded by lines with constant values of $v$ and $u$,
which in the Penrose diagram correspond to $45$ degrees lines.

On the left and right boundaries,
the time  coordinate $t$ diverges to $\pm \infty$ in the upper and lower sides, respectively.
From eqs. (\ref{Kruskalcoordinates}), a change of cutoff
from $\Lambda_1$ to $\Lambda_2$, implies a constant shift in the time coordinate
by $C \log \frac{\Lambda_2}{\Lambda_1}$.
For $\nu=1$ we recover the AdS asymptotic, $r^*(\infty)$ is finite and no shift is needed;
the Penrose diagram in this case is different and is the standard one of the BTZ black hole.

The BH has a left and a right boundary, where
two identical copies of a dual entangled WCFT live. 
To avoid divergences, the times at the left
and right boundaries  are evaluated
at the cutoff surface $r=\Lambda$, and 
are respectively denoted by $t_L$ and $t_R$.
If we take the two times going in opposite directions:
\beq
t_L \rightarrow t_L +\Delta t \, , \qquad t_R \rightarrow t_R  - \Delta t \, ,
\eeq
 the  entangled thermofield doublet is time-independent, because
this time shift corresponds to the time Killing vector of the BH solution.
If instead we take the two boundary times going in the same direction, i.e.
\beq
t_L \rightarrow t_L +\Delta t \, , \qquad t_R \rightarrow t_R  + \Delta t \, ,
\eeq
the BH solution is dual to a time-dependent thermofield doublet \cite{Hartman:2013qma}:
 \beq
|\Psi_{TFD} \rangle   \propto   \sum_n  e^{-E_n \beta/2- i E_n (t_L + t_R)} | E_{n} \rangle_R  | E_{n} \rangle_L \, .
\eeq
where $| E_{n} \rangle_{L,R}$ denotes the energy eigenstates of left and right boundary theories and
 $\beta$ is the inverse temperature.
Without loss of generality, we can choose 
\beq
t_L=t_R=\frac{t_b}{2} \,.
\eeq

\subsection{Non-rotating case}
\label{Sect-non-rotating}
\begin{figure}[h]
\begin{center}
\leavevmode
\epsfig{file=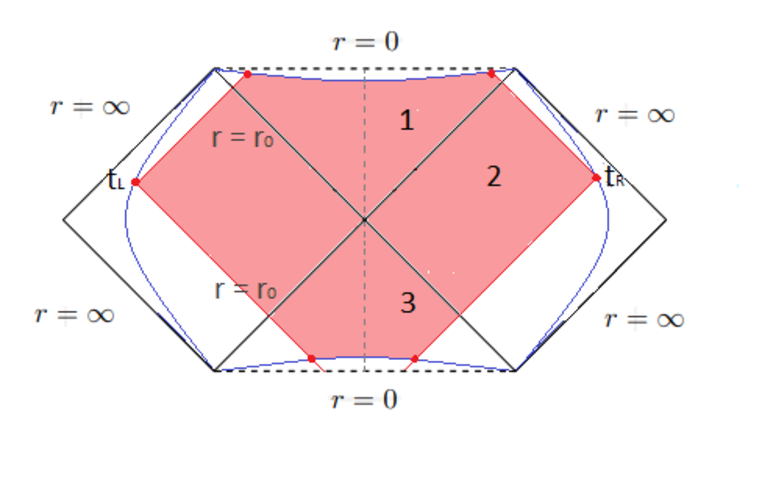, width =4.in}
\end{center}
\caption{Penrose diagram  for the non-rotating BH, with the WDW patch  for $t_b<t_C $.}
\label{fig1}
\end{figure}

The non-rotating case  corresponds to the values 
in eq. (\ref{non-rota});
for simplicity we  focus just on  $r_-=0$ and we  set $r_+=r_0$.
The analysis for the other value of $r_+/r_-$ in eq. (\ref{non-rota}) is analogous:
 it can be shown that it can be mapped
to $r_-=0$ by a change of variables  \cite{Jugeau:2010nq}.
The Penrose diagrams for the non-rotating case are shown in figures
\ref{fig1} and \ref{fig2}. 

The structure of the WDW patch in the non-rotating case changes with time; 
at early times it looks like in figure \ref{fig1}, while at late times like in
figure \ref{fig2}. In particular, there exists a critical time $ t_C $
 such that the bottom vertex of the patch touches the past singularity.
The critical time is given by
\beq
t_C = 2(r^*_{\Lambda} - r^* (0) )\, ,
\eeq
where $r^*_{\Lambda}=r^*(\Lambda)$.
We will separate the calculation of the action in  two cases.
At the end we will express the results
in terms of 
\beq
\tau=l (t_b - t_C) \, ,
\eeq 
where $\tau$ is the boundary time rescaled with curvature $l$
for dimensional purposes and with the origin translated 
at the critical time $t_C$.

\subsubsection{Initial times $ t_b<t_C $}

{\bf{ Bulk contributions:}}
We decompose the WDW patch into three regions and we use the symmetry of the configuration to write the bulk action as
\beq
I_{\mathcal{V}}  = 2 \le I^{1}_{\mathcal{V}} + I^{2}_{\mathcal{V}}  + I^{3}_{\mathcal{V}}  \ri \, ,
\eeq
where 
\bea
I^{1}_{\mathcal{V}}  &=& \frac{\mathcal{I}}{16 \pi G} \int_0^{2\pi} d \theta \int_{\varepsilon_0}^{r_0} dr  \int_0^{v - r^* (r)} dt = \frac{\mathcal{I}}{8 G} \int_{\varepsilon_0}^{r_0} dr \le \frac{t_b}{2} + r^*_{\Lambda} - r^* (r) \ri \, ,
\nl 
I^{2}_{\mathcal{V}}  &=& \frac{\mathcal{I}}{16 \pi G} \int_0^{2\pi} d \theta \int_{r_0}^{\Lambda} dr  \int_{u + r^* (r)}^{v - r^* (r)} dt = \frac{ \mathcal{I}}{4 G} \int_{r_0}^{\Lambda} dr \le  r^*_{\Lambda} - r^* (r) \ri \, ,
\nl
I^{3}_{\mathcal{V}}  &=&   \frac{\mathcal{I}}{16 \pi G} \int_0^{2\pi} d \theta \int_{\varepsilon_0}^{r_0} dr  \int_{u + r^* (r)}^0 dt = \frac{ \mathcal{I}}{8  G} \int_{\varepsilon_0}^{r_0} dr \le - \frac{t_b}{2} + r^*_{\Lambda} - r^* (r) \ri \, .
\eea
Summing all the contributions, we get the result
\beq
I_{\mathcal{V}}  =
\frac{\mathcal{I}}{2 G} \int_{\varepsilon_0}^{\Lambda} dr \le  r^*_{\Lambda} - r^* (r) \ri \equiv
I^0_{\mathcal{V}}  \, .
\eeq
This contribution is time-independent.

{\bf GHY surface  contributions:}
The constant $r$ surface, inside the horizon, is a spacelike surface 
whose induced metric in the $x^i=(t,\theta)$ coordinates reads: 
\beq
h_{ij} = l^2 \begin{pmatrix}
1 & \nu r \\
\nu r & \frac{r}{4} \Psi(r)
\end{pmatrix} \, , \qquad
\sqrt{h} = \frac{l^2}{2} \sqrt{(\nu^2 +3)r(r_0-r)} \, .
\eeq
The normal vector to these slices is
\beq
n^{\mu} = \le 0 \, , - \frac{1}{l} \sqrt{(\nu^2 +3)r(r_0-r)} \, , 0 \ri \, ,
\qquad n^\a n_\a = -1 \, ,
\label{normal-constant-r}
\eeq
and the extrinsic curvature is
\beq
K = \frac{1}{2 l} \sqrt{\nu^2 +3} \frac{2r - r_0}{\sqrt{r(r_0-r)} } \, .
\eeq
In the GHY we should then use $\varepsilon=-1$ because the surface is spacelike.
We are now able to compute the two contributions to the GHY term coming from the regions near the past and future singularities:
\beq
I^1_{\rm GHY} =
- \frac{ (\nu^2 +3) l}{16  G} \left[ \le 2r - r_0 \ri \le \frac{t_b}{2} +r^*_{\Lambda} - r^*(r) \ri  \right]_{r=\varepsilon_0} \, ,
\eeq
\beq
I^2_{\rm GHY}  =
- \frac{ (\nu^2 +3) l}{16 G} \left[ \le 2r - r_0 \ri \le - \frac{t_b}{2} +r^*_{\Lambda} - r^*(r) \ri  \right]_{r=\varepsilon_0} \, .
\eeq
The total GHY contribution then is:
\beq
I_{\rm GHY} = 2 \le I^1_{\rm GHY}+ I^2_{\rm GHY} \ri =
- \frac{ (\nu^2 +3) l}{4 G} \left[ \le 2r - r_0 \ri \le  r^*_{\Lambda} - r^*(r) \ri  \right]_{r=\varepsilon_0} \equiv I^0_{\rm GHY} \, ,
\eeq
which is time-independent.

{\bf Joint  contributions:}
There are four joints between null and 
spacelike surfaces at $r=\varepsilon_0$ (nearby  the future and past  singularities)
and two joints at $r=\Lambda$.
The normal to the constant $r$ spacelike surfaces is 
$n^\a$ given by eq. (\ref{normal-constant-r}), while the normal
to the lightlike surfaces are $u^\a$, $v^\a$ from eq. (\ref{null-normals}).
From eq. (\ref{jns}), the four joint contributions nearby the singularities vanish,
while the two joint contribution nearby the UV cutoff are time-independent
(see eq. \ref{jnn}).

{\bf Total:}
Summing all the terms coming from the bulk, the boundary and the joint contributions, we find
that the action of the WDW patch is time-independent.

\subsubsection{Later times $ t_b> t_C $}

\begin{figure}[h]
\begin{center}
\leavevmode
\epsfig{file=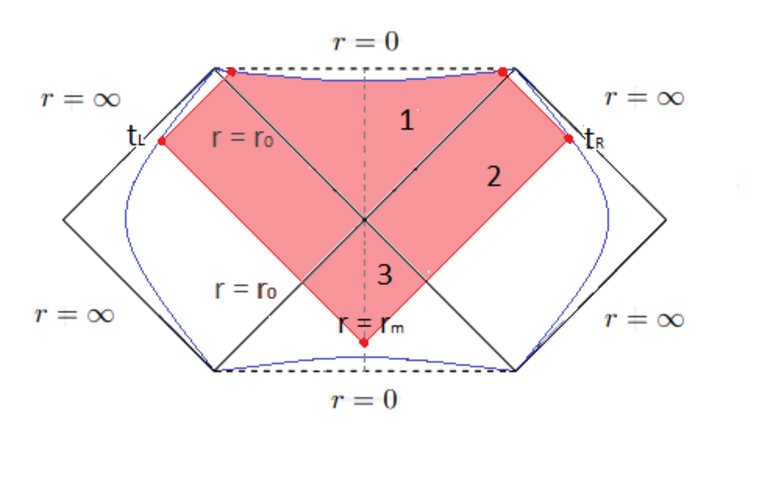, width =4.in}
\end{center}
\caption{Penrose diagram  for the non-rotating BH, with the WDW patch  for $t_b>t_C $.}
\label{fig2}
\end{figure}

After the critical time $ t_C , $ the WDW patch moves and the 
lower vertex of the diagram does not reach the past singularity (see figure \ref{fig2}).
This vertex is defined via the relation
\beq
\frac{t_b}{2} - r^*_{\Lambda} + r^* (r_m ) = 0 \, .
\label{definizione vertice basso}
\eeq
The evaluation of the null joint contributions will require the computation of the time derivative of the tortoise coordinate,
which is done by differentiating eq. (\ref{definizione vertice basso}):
\beq
\frac{d r_m}{dt_b} =
-\frac12 
\le \frac{dr^*(r_m)}{d r_m} \ri^{-1} \, .
\label{time-derivative}
\eeq

{\bf Bulk contributions:}
The bulk action is the same of the case $ t_b <t_C , $ apart from the last contribution which becomes
\beq
I^{3}_{\mathcal{V}}  (t_b >t_C) =
   \frac{\mathcal{I}}{16 \pi G} \int_0^{2\pi} d \theta \int_{r_m}^{r_0} dr  \int_{u + r^* (r)}^0 dt 
   = \frac{ \mathcal{I}}{8 G} \int_{r_m}^{r_0} dr \le - \frac{t_b}{2} + r^*_{\Lambda} - r^* (r) \ri \, .
\eeq
We can re-write this contribution in the following way:
\beq
 I^{3}_{\mathcal{V}}  (t_b >t_C)  = I^{3}_{\mathcal{V}}  (t_b <t_C) +
\frac{ \mathcal{I}}{8 G} \int^{r_m}_{\varepsilon_0} dr \le  \frac{t_b}{2} - r^*_{\Lambda} + r^* (r) \ri \, .
\eeq
Since the other contributions to the bulk action are unchanged, the total result is
\beq
I_{\mathcal{V}}  (t_b >t_C)  =
I^0_{\mathcal{V}}  + \frac{ \mathcal{I}}{4 G} \int^{r_m}_{\varepsilon_0} dr \le  \frac{t_b}{2} - r^*_{\Lambda} + r^* (r) \ri \, ,
\eeq
the first term being time-independent.
The time derivative of the bulk action then is:
\beq
\frac{dI_{\mathcal{V}} }{d t_b} (t_b >t_C)  =
 \frac{ \mathcal{I}}{8 G} \, r_m =
 \frac{1}{8 G} \left[ - \frac{l}{2} (\nu^2 +3) + \frac{\kappa c^2}{l} - \alpha a c \right] r_m \, ,
 \label{bulk-late-time-non-rot}
\eeq
where the defining relation (\ref{definizione vertice basso}) is used in order to obtain a vanishing
contribution from the upper integration extreme.

{\bf GHY surface  contributions:}
After the critical time $  t_C $ we only have a contribution from the future singularity,
 because the lower part of the WDW patch does not reach the past singularity.
We are only left with
\beq
I_{\rm GHY}  = 2 I^1_{\rm GHY}  =
- \frac{(\nu^2 +3) l}{8 G} \left[ \le 2r - r_0 \ri \le \frac{t_b}{2} +r^*_{\Lambda} - r^*(r) \ri  \right]_{r=\varepsilon_0} \, ,
\eeq
which is time-dependent.
The time derivative of this term gives:
\beq
\lim_{\varepsilon_0 \rightarrow 0} \frac{dI_{\rm GHY}}{d t_b} (t_b >t_C)  =
\frac{(\nu^2 +3)l}{16 G} \, r_0 \, .
\label{GHY-late-time}
\eeq

{\bf Joint  contributions:}
Following the same procedure of the case $ t_b <t_C$,  we find that the null joints 
at the UV cutoff give time-independent contributions, 
while the joint at the future singularity gives a vanishing result.
The contribution from the remaining  null-null  joint between $u^\a$ and  $v^\a$
at $ r=r_m$ is instead time-dependent, because $r_m$ is function of time (see eq. (\ref{time-derivative})).
We find that this contribution to the action is given by
eq. (\ref{jjoints}), with $\mathfrak{a}$ given by eq. (\ref{jnn}):
\beq
\mathfrak{a}= \log \left| A^2 \frac{ u^\a v_\a}{2}
 \right| = \log \left| A^2 \frac{1}{l^2} \frac{ \Psi(r) }{(\nu^2 +3) (r-r_0)} \right| \, .
 \label{null-joint}
\eeq
The normalization factor $A^2$ corresponds to an ambiguity
in the contribution to the action due to the null joint \cite{Lehner:2016vdi},
because the normalization of the two null normals $u^\a$ and  $v^\a$
which delimitate the WDW patch is in principle not fixed by the metric
 (see the discussion at the end of section \ref{sect-action}). 
The action contribution from eq. (\ref{null-joint}), evaluated for $ r=r_m$, gives: 
\beq
I_{\mathcal{J}} = - \frac{ l}{4 G} \sqrt{\frac{r_m}{4}
\Psi(r_m) } 
\log \left| \frac{l^2}{A^2} \frac{(\nu^2 +3)(r_m-r_0)}
{\Psi(r_m)}  \right| \, ,
\eeq
whose time derivatives is:
\beq
\begin{aligned}
 \frac{dI_{\mathcal{J}}}{d t_b}  = & - \frac{ l}{16 G}   \frac{d r_m}{dt_b}  
\frac{6(\nu^2 -1) r_m+ (\nu^2 +3) r_0 }{\sqrt{ r_m \left[ 3 (\nu^2 -1) r_m + (\nu^2 +3) r_0 \right]}}
\log \left| \frac{l^2}{A^2} \frac{(\nu^2 +3)(r_m-r_0)}{\Psi(r_m)}  \right|
+ \\
& - \frac{ l}{8 G}  \frac{d r_m}{dt_b}   
\frac{4 \nu^2  r_0 \sqrt{r_m \left[ 3 (\nu^2 -1) r_m + (\nu^2 +3) r_0 \right] }}{(r_m-r_0)\le 3 r_m (\nu^2 -1) + (\nu^2 +3) r_0 \ri}  \, .
\end{aligned}
\eeq
Inserting eq. (\ref{time-derivative}) we obtain:
\beq
\begin{aligned}
 \frac{dI_{\mathcal{J}}}{d t_b}  = &  \frac{l}{32 G}  
\frac{(\nu^2 +3)(r_m-r_0) \le 6(\nu^2 -1) r_m + (\nu^2 +3) r_0 \ri}{ 3 (\nu^2 -1) r_m + (\nu^2 +3) r_0 }
\log \left| \frac{l^2}{A^2} \frac{(\nu^2 +3)(r_m-r_0)}{\Psi(r_m)}  \right|
+ \\
& + \frac{l}{16 G}  \frac{4 \nu^2 (\nu^2 +3) r_m r_0 }{ 3 r_m (\nu^2 -1) + (\nu^2 +3) r_0}  \, .
\end{aligned}
\eeq

{\bf Total:}
The total time derivative of the action is finally given by
\beq
\begin{aligned}
\frac{d I}{d t_b} & = \frac{1}{8 G} \left[ - \frac{l}{2} (\nu^2 +3) + \frac{\kappa c^2}{l} 
- \a a c \right] r_m + \frac{(\nu^2 +3)l}{16 G} \, r_0  + \frac{l}{16 G}
  \frac{4 \nu^2 (\nu^2 +3) r_m r_0 }{ 3 r_m (\nu^2 -1) + (\nu^2 +3) r_0}  \\
& + \frac{l}{32 G}  
\frac{(\nu^2 +3)(r_m-r_0) \le 6(\nu^2 -1) r_m + (\nu^2 +3) r_0 \ri}{ 3 (\nu^2 -1) r_m + (\nu^2 +3) r_0 }
\log \left| \frac{l^2}{A^2} \frac{(\nu^2 +3)(r_m-r_0)}{\Psi(r_m)}  \right| \, .
\end{aligned}
\label{gran-totale1}
\eeq

We can now perform the late time limit of the previous rate. 
In this limit $ r_m \rightarrow r_0 , $ which implies that the term in the second line vanishes and we find:
\beq
\lim_{t_b \rightarrow \infty} \frac{d I}{d t_b} 
= \frac{(\nu^2 +3)l}{16 G} \, r_0 + \frac{1}{8 G} \le \frac{\kappa}{l} c^2 - \alpha  a c \ri r_0  \, .
\eeq
Note that the general result (\ref{gran-totale1}) depends on $A^2$, 
while its late time limit does not.
Using the value of $a$ given in eq. (\ref{a-action}),
we can now evaluate the combination appearing in the rate of the action
\beq
\frac{\kappa}{l} c^2 - \alpha a c = 0 \, .
\eeq
We finally obtain:
\beq
\lim_{t_b \rightarrow \infty} \frac{1}{l} \frac{d I}{d t_b} 
= \lim_{\tau \rightarrow \infty} \frac{d I}{d \tau} = \frac{\nu^2 +3}{16 G} \, r_0 = M = TS \, .
\eeq
This late-time results can also be recovered using the approach
by  \cite{Brown:2015lvg} (see Appendix \ref{another-way} for details).

\begin{figure}[h]
\begin{center}
\leavevmode
\epsfig{file=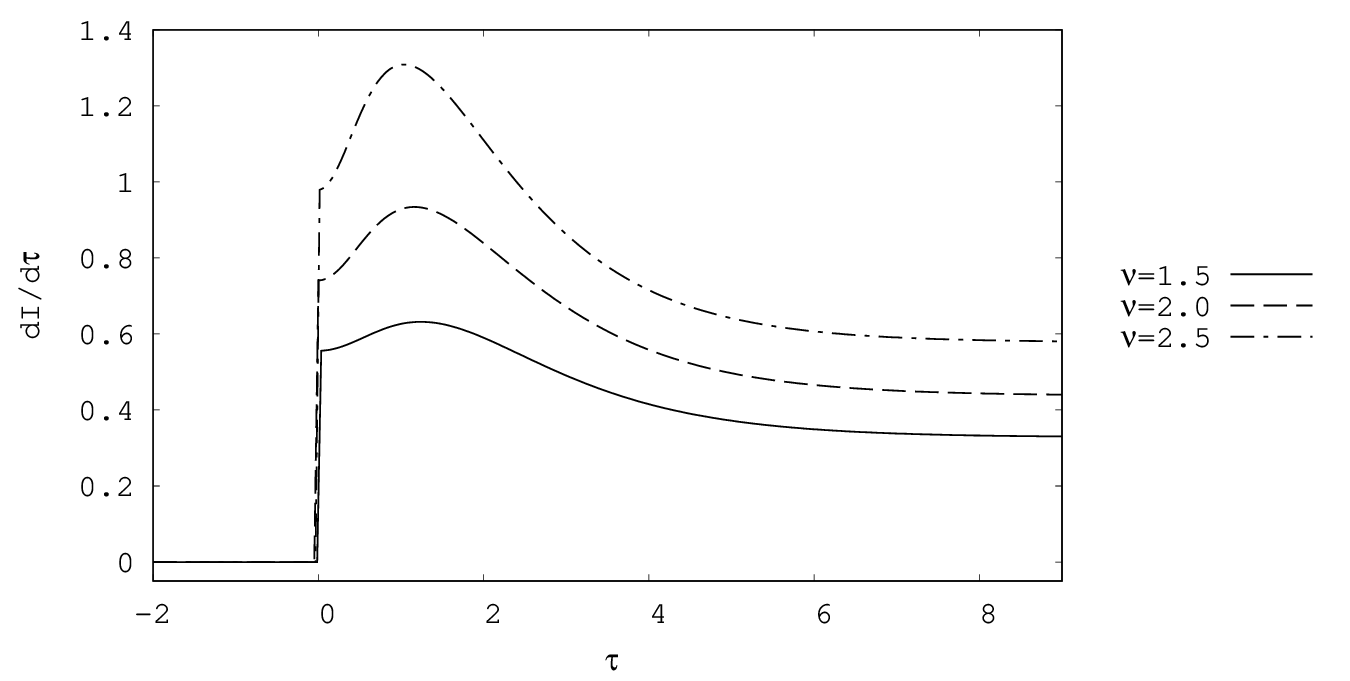, width =4.5in}
\end{center}
\caption{Time dependence of the WDW action in the non-rotating case
for different values of $\nu$. We set $G=1$, $l=1$, $r_0=1$ 
and $A=2$. The critical time $t_C$ corresponds to $\tau=0$. }
\label{CA-time-dep1}
\end{figure}
\begin{figure}[h]
\begin{center}
\leavevmode
\epsfig{file=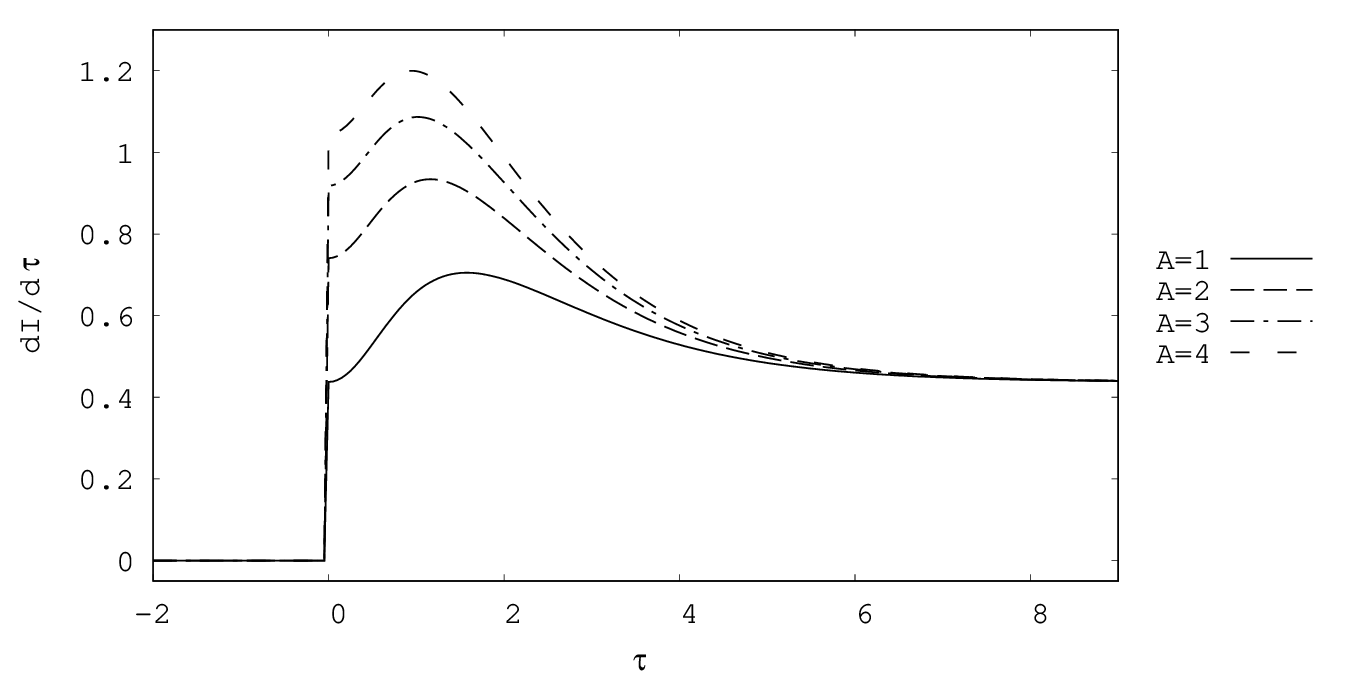, width =4.5in}
\end{center}
\caption{Time dependence of the WDW action in the non-rotating case for different values of 
the parameter $A$. We set $G=1$, $l=1$, $r_0=1$ 
and $\nu=2$.}
\label{CA-time-dep-2}
\end{figure}
Numerical plots of the time dependence of the action rate (\ref{gran-totale1})
 for different values of $\nu$ 
are shown in figure \ref{CA-time-dep1}.
The same qualitative structure as for the AdS case \cite{Carmi:2017jqz} is found;
in particular the growth rate of the action is a decreasing function at late times.
As in \cite{Carmi:2017jqz}, 
the late-time limit then overshoots the asymptotic rate, which 
was previously believed \cite{Brown:2015lvg}
to be associated to an universal
upper bound, conjectured by Lloyd \cite{Lloyd}. 
There is some dependence
at finite time on the parameter $A$, see figure \ref{CA-time-dep-2};
this is a feature also of the AdS case \cite{Lehner:2016vdi,Chapman:2016hwi,Carmi:2017jqz}. The late-time limit is instead independent from $A$.

 

\subsection{Rotating case}
\label{Sect-rotating}
\begin{figure}[h]
\begin{center}
\leavevmode
\epsfig{file=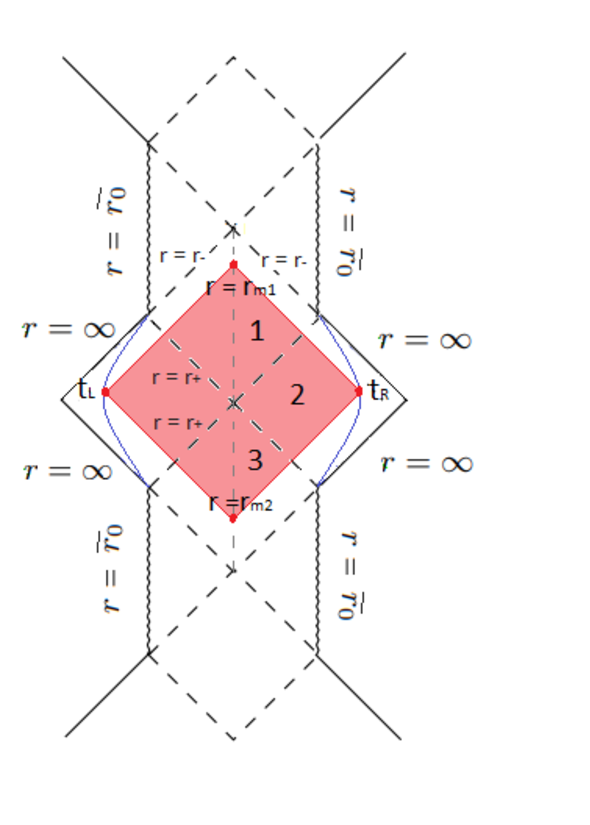, width =3.0in}
\end{center}
\caption{Penrose diagram for the WDW patch in the rotating case.}
\label{fig3}
\end{figure}

In the rotating case (see figure \ref{fig3}) we
 do not need to distinguish between initial and later times, 
 because in this case the form 
of the WDW patch is the same at any time and the complexity is already non-vanishing at initial times. We define $\tau=l \, t_b$.
We call $  r_{m1}, r_{m2} $ the null joints referring respectively to the top and bottom vertices of the spacetime region of interest.
Due to the structure of the Penrose diagram in the rotating case (similar to the 3+1 dimensional diagram 
for a Reissner-Nordstrom black hole), we do not have boundaries contributing to the GHY term.

The definition of the null joints in terms of the tortoise coordinates are:
\beq
\frac{t_b}{2} + r^*_{\Lambda} - r^* (r_{m1} ) = 0 \, , \qquad
\frac{t_b}{2} - r^*_{\Lambda} + r^* (r_{m2} ) = 0 \, .
\label{definizione dei joint}
\eeq
It will be useful to differentiate with respect to time these expressions to find
\beq
\frac{d r_{m1}}{dt_b} =
\frac12 \le \frac{dr^*}{d r_{m1}} \ri^{-1} \, , \qquad
\frac{d r_{m2}}{dt_b} =
-\frac12 \le \frac{dr^*}{d r_{m2}} \ri^{-1} \, .
\label{derivate joint}
\eeq

{\bf Bulk contributions:}
We can still split the WDW patch into three regions covering only the right half of the diagram, which contribute as
\bea
I^{1}_{\mathcal{V}}  &=&  \frac{ \mathcal{I}}{8 G} \int_{r_{m1}}^{r_+} dr \le \frac{t_b}{2} + r^*_{\Lambda} - r^* (r) \ri \, ,
\qquad
I^{2}_{\mathcal{V}}  =  \frac{ \mathcal{I}}{4 G} \int_{r_+}^{\Lambda} dr \le  r^*_{\Lambda} - r^* (r) \ri \, , \nl 
I^{3}_{\mathcal{V}}  &=&   \frac{ \mathcal{I}}{8 G} \int_{r_{m2}}^{r_+} dr \le - \frac{t_b}{2} + r^*_{\Lambda} - r^* (r) \ri \, .
\eea
The whole bulk contribution then amounts to
\bea
I_{\mathcal{V}}  &=& \frac{ \mathcal{I}}{2 G} \int_{r_+}^{\Lambda} dr \le  r^*_{\Lambda} - r^* (r) \ri  + \nl
&+& \frac{ \mathcal{I}}{4 G} \left[ \int_{r_{m1}}^{r_+} dr \le \frac{t_b}{2} + r^*_{\Lambda} - r^* (r) \ri 
+ \int^{r_{m2}}_{r_+} dr \le \frac{t_b}{2} - r^*_{\Lambda} + r^* (r) \ri \right] \, .
\eea
The rate of the bulk action is
\beq
\frac{dI_{\mathcal{V}} }{d t_b}  =
 \frac{ \mathcal{I}}{8 G} (r_{m2}- r_{m1} ) \, ,
\eeq
where the relations (\ref{definizione dei joint}) are used to obtain a vanishing result when differentiating the ends of integration.
The result simplifies when performing the late time limit, when
$ r_{m1} \rightarrow r_+$ and  $ r_{m2} \rightarrow r_-$,
and the bulk action time-derivative becomes
\beq
\lim_{t_b \rightarrow \infty} \frac{d I_{\mathcal{V}} }{d t_b} = - \frac{(\nu^2 +3)l}{16 G} \, (r_+ - r_-) + \frac{1}{8 G} \le \frac{\kappa}{l} c^2 - \alpha  a c \ri (r_+ - r_-) \, .
\label{bulk-late-time-rot}
\eeq

{\bf{Null joint contributions}}:
As in the non-rotating case,  the joints at $ r= \Lambda $ give a time-independent contribution, 
and then they are not of interest to find the rate of complexity.
We have two time-dependent  contributions coming from the top and bottom joints.

As a function of $r$, these contributions are proportional to:
\beq
\mathfrak{a} = \eta \log \left| A^2 \frac12 u^\a v_\a \right|  = 
\eta\log \left| \frac{A^2}{l^2} \frac{r \Psi (r)}{(\nu^2 +3)(r-r_-)(r-r_+)} \right|
 \, .
\eeq
For  both $ r=r_{m1}$ and  $ r=r_{m2}$  we have to insert 
 $ \eta_1=\eta_2=1$.

The action of each joint then is: 
\beq
I_{\mathcal{J}}^{k} = - \frac{l}{4 G} \sqrt{\frac{r_{k}}{4} \Psi(r_{k})}
 \log \left| \frac{l^2}{A^2} F(r_k) \right| \, ,
 \qquad 
 F(r_k) \equiv \frac{(\nu^2 +3)(r_{k}-r_-)(r_{k}-r_+)}{r_{k} \Psi (r_{k})}  \, ,
\eeq
and $r_1=r_{m1}$,  $r_2=r_{m2}$.
We differentiate with respect to time the null joint contributions:
\bea
& & \frac{dI_{\mathcal{J}}^k}{d t_b}  =
   - \frac{ l}{8 G}   \frac{d r_{k}}{dt_b}   \left\{  
    \sqrt{r_{k} \Psi (r_{k})} 
\frac{d}{d r_{k}} \le 
\log \left| \frac{l^2}{A^2}   F(r_{k}) 
  \right| \ri 
+  \right. \nl
& & \left.   
 + \frac12 \frac{ 6(\nu^2 -1) r_{k} + (\nu^2 +3) (r_+ + r_-) -4 \nu \sqrt{(\nu^2 +3) r_+ r_-} }{\sqrt{ r_{k} \Psi(r_{k})}} 
\log \left| \frac{l^2}{A^2} F(r_{k})    \right|
   \right\} \, .
\eea
Using eqs. (\ref{derivate joint}) in the previous expression,
 it is possible to find the complete time dependence of the null contributions.
 In the late-time limit, we find:
\beq
\lim_{t_b \rightarrow \infty} \frac{d I_{\mathcal{J}}^k}{d t_b} =  \frac{(\nu^2 +3)l}{16 G} \, (r_+ - r_-) \, ,
\qquad
k=1,2 \, .
\eeq

{\bf Total:}
Summing all the previous asymptotic expressions, the late-time limit of the action growth is:
\beq
\lim_{t_b \rightarrow \infty} \frac{d I}{d t_b} = \frac{(\nu^2 +3)l}{16 G} \, (r_+ - r_-) - \frac{1}{8 G} \le \frac{\kappa}{l} c^2 - \alpha  a c \ri (r_+ - r_-) \, .
\eeq
Taking into account eq. (\ref{a-action}) we finally find: 
\beq
\lim_{t_b \rightarrow \infty} \frac{1}{l} \frac{d I}{d t_b} = \lim_{\tau \rightarrow \infty} \frac{d I}{d \tau} = \frac{(\nu^2 +3)}{16 G} \, (r_+ - r_-)  = TS \, .
\eeq
The late-time limit can be recovered also with the
methods introduced in  \cite{Brown:2015lvg} and the results agree; 
details of the explicit calculation  can be found  in appendix \ref{another-way}.

\section{Conclusions}
\label{sect-conclu}
In this paper we investigated the CA conjecture
for WAdS BHs realized as solutions of Einstein gravity plus matter.
We have found that, both in the rotating and in the non-rotating cases, the asymptotic limit
of the action in the WDW patch is:
\beq 
 \lim_{\tau \rightarrow \infty} \frac{d I}{d \tau}= TS \, ,
\qquad
T S = \frac{ (r_+ -r_-)(3+\nu^2)}{16 G} \, .
\eeq
In the rotating case, the only terms which contribute are the bulk and the joints term,
while in the non-rotating case there is also a surface GHY contribution.
Although the details of the calculation are 
 quite different,  the final result is a continuous function of the parameters
 of the solution $(r_+,r_-)$.
A curious feature of  the non-rotating case is that there exists an initial time period
($t<t_c$)  in which complexity is constant; this is the same as in the AdS 
case  \cite{Carmi:2017jqz}.

The results can be compared to the ones from the CV conjecture,
studied in \cite{Auzzi:2018zdu}:
\beq
 \lim_{\tau \rightarrow \infty}
\frac{d V}{d \tau} = \frac{\pi l}{2}
(r_+ -r_-)  \sqrt{3+\nu^2}  = TS \,  \frac{8 \pi G l }{\sqrt{3+\nu^2}} \, .
\eeq
Already in the AdS case the CA conjecture is known
to be more universal, because no explicit factor of the curvature $l$
related to the asymptotic of the spacetime is needed.
In the case of WAdS, this behavior is confirmed:
the CA gives as a result $TS$, independently from the two parameters
$(l,\nu)$ which determine the space-time asymptotic, while in the CV
a factor $\frac{\sqrt{3+\nu^2}}{8 \pi G l}$ 
should be inserted in front of the volume
in order to match with the CA.

WAdS BHs can be realized also as solutions 
of TMG (Topological Massive Gravity) and NMG (New Massive Gravity).
It would be interesting to study both CA and CV in these examples, in order
to get control on both the conjectures in the case of higher derivatives terms in
the gravity action.  The CA conjecture for higher derivatives gravity
was already  studied by several authors in
in \cite{Alishahiha:2017hwg,Guo:2017rul,Ghodrati:2017roz,Qaemmaqami:2017lzs},
but always in the late-time limit.  In particular, ref. \cite{Ghodrati:2017roz} studied the late-time limit
of CA conjecture for WAdS BHs in TMG; the asymptotic growth of the action
is not proportional to $TS$.

Another important open problem is to study complexity from the 
field theory dual. In particular, it would be interesting to  generalize
the Liouville action  \cite{Caputa:2017yrh,Bhattacharyya:2018wym} approach to WAdS.

\section*{Appendix}
\addtocontents{toc}{\protect\setcounter{tocdepth}{1}}
\appendix

\section{Comparison with ref. \cite{Banados:2005da}.}
\label{comparison}

Let us fix the couplings $(\kappa,L,\a)$ in the action (\ref{S-bulk}); 
the field equation determine the solution parameters $(\nu,l)$ as follows:
\beq
\nu^2 + \frac{2 	\kappa}{\a L^2} \nu -3=0 \, , \qquad \nu = -\frac{\kappa}{\a L^2} \pm \sqrt{ \frac{\kappa^2}{\a^2 L^4} +3} \, ,
\qquad l=\frac{\kappa \nu}{\a} \, .
\eeq
 Note that the following transformation on the couplings and fields
 gives an invariance of the action (\ref{S-bulk}):
\beq
\kappa \rightarrow -\kappa \, , \qquad \a \rightarrow -\a \, , \qquad A_\mu \rightarrow  i A_\mu \, .
\label{fsymmetry}
\eeq
This is just a formal trick, because the gauge field becomes imaginary.
This is useful in order to match with the results of  \cite{Banados:2005da},
because they consider just the $\kappa=1$ case.

The metric used in \cite{Banados:2005da} reads: 
\beq
ds^2 = p d\tilde{t}^2 + \frac{d\tilde{r}^2}{h^2-pq} + 2 h d\tilde{t} d\tilde{\theta} + q d \tilde{\theta}^2 \, ,
\label{metric-BBCG}
\eeq
 We can put the metric (\ref{BHole}) in the form (\ref{metric-BBCG}) by means of the coordinate transformations:
\beq
\tilde{t}  = \sqrt{\frac{l^3}{\omega}} t  \, , \qquad
\tilde{r}=r - \frac{\sqrt{r_+ r_- (\nu^2 +3)}}{2 \nu} \, , \qquad
\tilde{\theta} = -  \frac{\sqrt{\omega l^3}}{2} \theta \, ,
\eeq
where
\beq
\omega = \frac{\nu^2 +3}{2 l} \le  (r_+ + r_-) - \frac{\sqrt{r_+ r_- (\nu^2 +3)}}{\nu} \ri \, .
\eeq
 Let us introduce
  \bea
 \gamma^2 &=& 
 \frac{l}{\omega}  \frac{3(1-\nu^2)}{3-\nu^2}\, , 
 \qquad
 \mu = \frac{\omega}{8 G l} \, , 
 \nl 
4 G \mathcal{J} &=& (-\kappa) \frac{ 2 \nu  (r_+ + r_-) \sqrt{r_+ r_- (\nu^2 +3)} - (5\nu^2 +3)r_+ r_-}  
{2 l  \le \nu  (r_+ + r_-)  - \sqrt{r_+ r_- (\nu^2 +3)}  \ri} \, .
\eea
The quantity $\gamma^2$ is negative for $\nu>1$.
The functions appearing in the metric (\ref{metric-BBCG}) then are:
\beq
p(\tilde{r}) = 8 G \mu  \, , \qquad 
h (\tilde{r})  =-2 \frac{\nu}{l} \tilde{r}  \, ,
 \qquad
q (\tilde{r}) =  - 2 \frac{\gamma^2}{L^2} \tilde{r}^2
+2 \tilde{r}
-\frac{4G \mathcal{J} }{\a}  \, .
\eeq
Only the linear part in $ \tilde{r}$ of 
the $U(1)$ gauge field $A_{\tilde{\theta}}$ is determined by the equations of motion:
\beq
A_{\tilde{\theta}} (\tilde{r})= E \mp \frac{2\gamma}{L \sqrt{\kappa} } \tilde{r} \, ,
\label{ggau}
\eeq
The constant part,  denoted by $E$, does not enter both
the equations of motion and the calculation of the action, so we ignore it.
Moreover, the $\mp$ sign in eq. (\ref{ggau}) should be taken 
in correspondence of the $\pm$  sign of the second equation in (\ref{gaugau}).

The constant value of  $A_{\tilde{t}} $ is not determined by the equations of motion,
but affects the value of the bulk part of the action.
In the $\kappa=1$ case, it can be extracted from \cite{Banados:2005da}:
\beq
A_{\tilde{t}} (\tilde{r}) =  \frac{{\a} ^2 L^2 -1}{\gamma {\a} L} + \zeta \,  ,\qquad
A_t=\frac{d \tilde{t}}{d t} A_{\tilde{t}}=
-\frac{l}{\nu} \sqrt{\frac32} \sqrt{1-\nu^2} +\zeta \sqrt{\frac{l^3}{\omega}} \, .
\eeq
The value of $\zeta$ affects the way in which the physical mass is associated
to the Killing vector $\p/\p t$; gauge invariance of the result is recovered by $\zeta=0$.
For $\kappa=-1$, we can use the symmetry (\ref{fsymmetry})
to match with \cite{Banados:2005da}.
This gives the gauge field  $A_t$:
\beq
A_t=a=
\frac{l}{\nu} \sqrt{\frac32} \sqrt{\nu^2-1} 
+\zeta \sqrt{\frac{l^3}{\omega}}
 \, ,
\eeq
which reduces to  eq. (\ref{a-action}) for $\zeta=0$.

\section{Another way to compute the asymptotic growth of action}
\label{another-way}

The asymptotic growth of the action of the WDW patch can be computed 
also in the way introduced in \cite{Brown:2015lvg}.
This is a cross-check of our calculation.

 \begin{figure}[h]
\begin{center}
\leavevmode
\epsfig{file=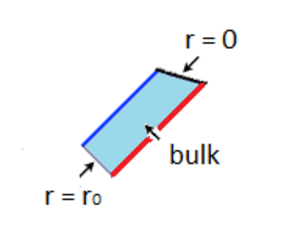, width =1.5in}
\end{center}
\caption{Asymptotic contributions for the non-rotating case.}
\label{SUS1}
\end{figure}
{\bf Non-rotating case:}
The relevant region in the WDW patch is shown in figure \ref{SUS1}.
The time derivative of the bulk contribution is given by
(\ref{bulk-late-time-non-rot}).
The time derivative of the GHY term nearby the singularity 
is given by eq. (\ref{GHY-late-time}).
The contribution from the joint at $r=r_m$ is replaced 
 by the GHY term nearby the horizon:
 \beq
\Delta I_{\rm GHY}^{r_0}  = 
\frac{(\nu^2 +3) l}{16 G}  \Delta t_b    \left[ 2r - r_0  \right]_{r=r_0} \, ,
\eeq
 which in the asymptotic limit gives the same contribution as the null joint.

 \begin{figure}[h]
\begin{center}
\leavevmode
\epsfig{file=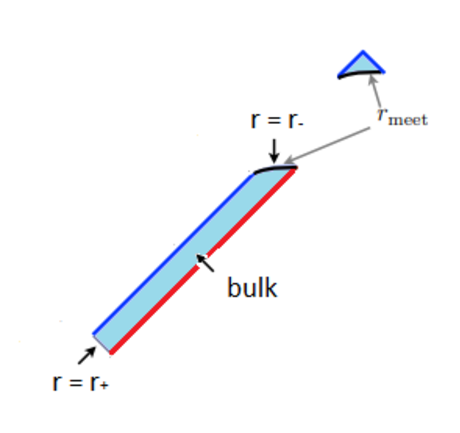, width =2.5in}
\end{center}
\caption{Asymptotic contributions for the rotating case.}
\label{SUS2}
\end{figure}
{\bf Rotating case:} 
The region is depicted in figure \ref{SUS2}.
 The bulk contribution is still given by eq. (\ref{bulk-late-time-rot}).
 The two null joints contributions are replaced by the GHY term evaluated
 on two constant-$r$ surfaces, one at $r \approx r_-$ and one at $r \approx r_+$.
 The induced metric on these constant-$r$ surfaces is:
 \beq
h_{ij} = l^2 \begin{pmatrix}
1 & \nu r -\frac12 \sqrt{(3+\nu^2) r_+ r_-} \\
\nu r -\frac12 \sqrt{(3+\nu^2) r_+ r_-} & \frac{r}{4} \Psi(r)
\end{pmatrix} \, , 
\eeq
\beq
\sqrt{h} = \frac{l^2}{2} \sqrt{(\nu^2 +3) (r_+-r) (r-r_-)} \, .
\eeq
 The normal vector to these slices is
\beq
n^{\mu} = \le 0 \, , - \frac{1}{l} \sqrt{(\nu^2 +3)(r_+  -  r)( r- r_- )} \, , 0 \ri \, ,
\qquad n^\a n_\a = -1 \, ,
\eeq
and the extrinsic curvature is
\beq
K = \frac{  \sqrt{\nu^2 +3} }{2 l} \,  \frac{2r - r_+ - r_-}{\sqrt{  (r_+  -  r)( r- r_- )  } } \, .
\eeq
 The GHY term nearby the inner horizon gives:
 \beq
 \frac{dI^{r_-}_{\rm GHY}}{d t_b} =-\frac{l}{4} \sqrt{\nu^2+3} \left[ 2 r -r_+ -r_- \right]_{r=r_-} \, ,
 \eeq
 while the term from the outer horizon
  \beq
 \frac{dI^{r_+}_{\rm GHY}}{d t_b} =\frac{l}{4} \sqrt{\nu^2+3} \left[ 2 r -r_+ -r_- \right]_{r=r_+} \, .
 \eeq
 These two contributions give the same result as the asymptotic contributions from the joints.


\begin{thebibliography}{1}

\bibitem{Ryu:2006bv}
  S.~Ryu and T.~Takayanagi,
  Phys.\ Rev.\ Lett.\  {\bf 96} (2006) 181602
  doi:10.1103/PhysRevLett.96.181602
  [hep-th/0603001].

\bibitem{Casini:2011kv}
  H.~Casini, M.~Huerta and R.~C.~Myers,
  JHEP {\bf 1105} (2011) 036
  doi:10.1007/JHEP05(2011)036
  [arXiv:1102.0440 [hep-th]].


\bibitem{Lewkowycz:2013nqa}
  A.~Lewkowycz and J.~Maldacena,
  JHEP {\bf 1308} (2013) 090
  doi:10.1007/JHEP08(2013)090
  [arXiv:1304.4926 [hep-th]].


\bibitem{Bekenstein:1973ur}
  J.~D.~Bekenstein,
  Phys.\ Rev.\ D {\bf 7} (1973) 2333.
  doi:10.1103/PhysRevD.7.2333




\bibitem{Bardeen:1973gs}
  J.~M.~Bardeen, B.~Carter and S.~W.~Hawking,
  Commun.\ Math.\ Phys.\  {\bf 31} (1973) 161.
  doi:10.1007/BF01645742
  
 
\bibitem{Maldacena:2001kr}
  J.~M.~Maldacena,
  JHEP {\bf 0304} (2003) 021
  doi:10.1088/1126-6708/2003/04/021
  [hep-th/0106112].

\bibitem{Hartman:2013qma}
  T.~Hartman and J.~Maldacena,
  JHEP {\bf 1305} (2013) 014
  doi:10.1007/JHEP05(2013)014
  [arXiv:1303.1080 [hep-th]].

\bibitem{Susskind:2014rva}
  L.~Susskind,
  [Fortsch.\ Phys.\  {\bf 64} (2016) 24]
   Addendum: Fortsch.\ Phys.\  {\bf 64} (2016) 44
  doi:10.1002/prop.201500093, 10.1002/prop.201500092
  [arXiv:1403.5695 [hep-th], arXiv:1402.5674 [hep-th]].


\bibitem{Susskind:2014moa}
  L.~Susskind,
  Fortsch.\ Phys.\  {\bf 64} (2016) 49
  doi:10.1002/prop.201500095
  [arXiv:1411.0690 [hep-th]].


\bibitem{Jefferson:2017sdb}
  R.~Jefferson and R.~C.~Myers,
  JHEP {\bf 1710} (2017) 107
  doi:10.1007/JHEP10(2017)107
  [arXiv:1707.08570 [hep-th]].

\bibitem{Chapman:2017rqy}
  S.~Chapman, M.~P.~Heller, H.~Marrochio and F.~Pastawski,
  Phys.\ Rev.\ Lett.\  {\bf 120} (2018) no.12,  121602
  doi:10.1103/PhysRevLett.120.121602
  [arXiv:1707.08582 [hep-th]].

\bibitem{Hashimoto:2017fga}
  K.~Hashimoto, N.~Iizuka and S.~Sugishita,
  Phys.\ Rev.\ D {\bf 96} (2017) no.12,  126001
  doi:10.1103/PhysRevD.96.126001
  [arXiv:1707.03840 [hep-th]].
  
\bibitem{Kim:2017qrq}
  R.~Q.~Yang, C.~Niu, C.~Y.~Zhang and K.~Y.~Kim,
  JHEP {\bf 1802} (2018) 082
  doi:10.1007/JHEP02(2018)082
  [arXiv:1710.00600 [hep-th]].
  
  \bibitem{Khan:2018rzm}
  R.~Khan, C.~Krishnan and S.~Sharma,
  arXiv:1801.07620 [hep-th].

\bibitem{Hackl:2018ptj}
  L.~Hackl and R.~C.~Myers,
  arXiv:1803.10638 [hep-th].


\bibitem{Caputa:2017yrh}
  P.~Caputa, N.~Kundu, M.~Miyaji, T.~Takayanagi and K.~Watanabe,
  JHEP {\bf 1711} (2017) 097
  doi:10.1007/JHEP11(2017)097
  [arXiv:1706.07056 [hep-th]].
  
\bibitem{Bhattacharyya:2018wym}
  A.~Bhattacharyya, P.~Caputa, S.~R.~Das, N.~Kundu, M.~Miyaji and T.~Takayanagi,
  arXiv:1804.01999 [hep-th].
  

  
\bibitem{Swingle:2009bg}
  B.~Swingle,
  Phys.\ Rev.\ D {\bf 86} (2012) 065007
  doi:10.1103/PhysRevD.86.065007
  [arXiv:0905.1317 [cond-mat.str-el]].

 \bibitem{Yang:2018nda}
  R.~Q.~Yang, Y.~S.~An, C.~Niu, C.~Y.~Zhang and K.~Y.~Kim,
  arXiv:1803.01797 [hep-th].
  
\bibitem{Hashimoto:2018bmb}
  K.~Hashimoto, N.~Iizuka and S.~Sugishita,
  arXiv:1805.04226 [hep-th].

\bibitem{Stanford:2014jda}
  D.~Stanford and L.~Susskind,
  Phys.\ Rev.\ D {\bf 90} (2014) no.12,  126007
  doi:10.1103/PhysRevD.90.126007
  [arXiv:1406.2678 [hep-th]].

\bibitem{Brown:2015bva}
  A.~R.~Brown, D.~A.~Roberts, L.~Susskind, B.~Swingle and Y.~Zhao,
  Phys.\ Rev.\ Lett.\  {\bf 116} (2016) no.19,  191301
  doi:10.1103/PhysRevLett.116.191301
  [arXiv:1509.07876 [hep-th]].

\bibitem{Brown:2015lvg}
  A.~R.~Brown, D.~A.~Roberts, L.~Susskind, B.~Swingle and Y.~Zhao,
  Phys.\ Rev.\ D {\bf 93} (2016) no.8,  086006
  doi:10.1103/PhysRevD.93.086006
  [arXiv:1512.04993 [hep-th]].

\bibitem{Hayward:1993my}
  G.~Hayward,
  Phys.\ Rev.\ D {\bf 47} (1993) 3275.
  doi:10.1103/PhysRevD.47.3275
  

\bibitem{Lehner:2016vdi}
  L.~Lehner, R.~C.~Myers, E.~Poisson and R.~D.~Sorkin,
  Phys.\ Rev.\ D {\bf 94} (2016) no.8,  084046
  doi:10.1103/PhysRevD.94.084046
  [arXiv:1609.00207 [hep-th]].
  


\bibitem{Chapman:2016hwi}
  S.~Chapman, H.~Marrochio and R.~C.~Myers,
  JHEP {\bf 1701} (2017) 062
  doi:10.1007/JHEP01(2017)062
  [arXiv:1610.08063 [hep-th]].

\bibitem{Carmi:2017jqz}
  D.~Carmi, S.~Chapman, H.~Marrochio, R.~C.~Myers and S.~Sugishita,
  JHEP {\bf 1711} (2017) 188
  doi:10.1007/JHEP11(2017)188
  [arXiv:1709.10184 [hep-th]].

\bibitem{Neiman:2012fx}
  Y.~Neiman,
  arXiv:1212.2922 [hep-th].

\bibitem{Parattu:2015gga}
  K.~Parattu, S.~Chakraborty, B.~R.~Majhi and T.~Padmanabhan,
  Gen.\ Rel.\ Grav.\  {\bf 48} (2016) no.7,  94
  doi:10.1007/s10714-016-2093-7
  [arXiv:1501.01053 [gr-qc]].
  
\bibitem{Cai:2016xho}
  R.~G.~Cai, S.~M.~Ruan, S.~J.~Wang, R.~Q.~Yang and R.~H.~Peng,
  JHEP {\bf 1609} (2016) 161
  doi:10.1007/JHEP09(2016)161
  [arXiv:1606.08307 [gr-qc]].
  
\bibitem{Barbon:2015ria}
  J.~L.~F.~Barbon and E.~Rabinovici,
  JHEP {\bf 1601} (2016) 084
  doi:10.1007/JHEP01(2016)084
  [arXiv:1509.09291 [hep-th]].

\bibitem{Bolognesi:2018ion}
  S.~Bolognesi, E.~Rabinovici and S.~R.~Roy,
  arXiv:1802.02045 [hep-th].

\bibitem{Reynolds:2017jfs}
  A.~P.~Reynolds and S.~F.~Ross,
  Class.\ Quant.\ Grav.\  {\bf 35} (2018) no.9,  095006
  doi:10.1088/1361-6382/aab32d
  [arXiv:1712.03732 [hep-th]].
  
  \bibitem{Moosa:2017yvt}
  M.~Moosa,
  JHEP {\bf 1803} (2018) 031
  doi:10.1007/JHEP03(2018)031
  [arXiv:1711.02668 [hep-th]].

\bibitem{Moosa:2017yiz}
  M.~Moosa,
  Phys.\ Rev.\ D {\bf 97} (2018) no.10,  106016
  doi:10.1103/PhysRevD.97.106016
  [arXiv:1712.07137 [hep-th]].

\bibitem{Chapman:2018dem}
  S.~Chapman, H.~Marrochio and R.~C.~Myers,
  arXiv:1804.07410 [hep-th].
  
\bibitem{Chapman:2018lsv}
  S.~Chapman, H.~Marrochio and R.~C.~Myers,
  arXiv:1805.07262 [hep-th].
  
\bibitem{Swingle:2017zcd}
  B.~Swingle and Y.~Wang,
  arXiv:1712.09826 [hep-th].
  
  
 \bibitem{An:2018xhv}
  Y.~S.~An and R.~H.~Peng,
  Phys.\ Rev.\ D {\bf 97} (2018) no.6,  066022
  doi:10.1103/PhysRevD.97.066022
  [arXiv:1801.03638 [hep-th]].




\bibitem{Anninos:2008fx}
  D.~Anninos, W.~Li, M.~Padi, W.~Song and A.~Strominger,
  JHEP {\bf 0903} (2009) 130
  doi:10.1088/1126-6708/2009/03/130
  [arXiv:0807.3040 [hep-th]].


\bibitem{Detournay:2012pc}
  S.~Detournay, T.~Hartman and D.~M.~Hofman,
  Phys.\ Rev.\ D {\bf 86} (2012) 124018
  doi:10.1103/PhysRevD.86.124018
  [arXiv:1210.0539 [hep-th]].

\bibitem{Hofman:2014loa}
  D.~M.~Hofman and B.~Rollier,
  Nucl.\ Phys.\ B {\bf 897} (2015) 1
  doi:10.1016/j.nuclphysb.2015.05.011
  [arXiv:1411.0672 [hep-th]].

\bibitem{Jensen:2017tnb}
  K.~Jensen,
  JHEP {\bf 1712} (2017) 111
  doi:10.1007/JHEP12(2017)111
  [arXiv:1710.11626 [hep-th]].

\bibitem{Anninos:2013nja}
  D.~Anninos, J.~Samani and E.~Shaghoulian,
  JHEP {\bf 1402} (2014) 118
  doi:10.1007/JHEP02(2014)118
  [arXiv:1309.2579 [hep-th]].

\bibitem{Castro:2015csg}
  A.~Castro, D.~M.~Hofman and N.~Iqbal,
  JHEP {\bf 1602} (2016) 033
  doi:10.1007/JHEP02(2016)033
  [arXiv:1511.00707 [hep-th]].

\bibitem{Azeyanagi:2018har}
  T.~Azeyanagi, S.~Detournay and M.~Riegler,
  arXiv:1801.07263 [hep-th].

\bibitem{Song:2016pwx}
  W.~Song, Q.~Wen and J.~Xu,
  Phys.\ Rev.\ Lett.\  {\bf 117} (2016) no.1,  011602
  doi:10.1103/PhysRevLett.117.011602
  [arXiv:1601.02634 [hep-th]].



\bibitem{Song:2016gtd}
  W.~Song, Q.~Wen and J.~Xu,
  JHEP {\bf 1702} (2017) 067
  doi:10.1007/JHEP02(2017)067
  [arXiv:1610.00727 [hep-th]].

\bibitem{Auzzi:2018zdu}
 R.~Auzzi, S.~Baiguera and G.~Nardelli,
  JHEP {\bf 1806} (2018) 063
  doi:10.1007/JHEP06(2018)063
  [arXiv:1804.07521 [hep-th]].


\bibitem{Banados:2005da}
  M.~Banados, G.~Barnich, G.~Compere and A.~Gomberoff,
  Phys.\ Rev.\ D {\bf 73} (2006) 044006
  doi:10.1103/PhysRevD.73.044006
  [hep-th/0512105].


\bibitem{Moussa:2003fc}
  K.~A.~Moussa, G.~Clement and C.~Leygnac,
  Class.\ Quant.\ Grav.\  {\bf 20} (2003) L277
  doi:10.1088/0264-9381/20/24/L01
  [gr-qc/0303042].


\bibitem{Bouchareb:2007yx}
  A.~Bouchareb and G.~Clement,
  Class.\ Quant.\ Grav.\  {\bf 24} (2007) 5581
  doi:10.1088/0264-9381/24/22/018
  [arXiv:0706.0263 [gr-qc]].




\bibitem{Banados:1992wn}
  M.~Banados, C.~Teitelboim and J.~Zanelli,
  Phys.\ Rev.\ Lett.\  {\bf 69} (1992) 1849
  doi:10.1103/PhysRevLett.69.1849
  [hep-th/9204099].
  
\bibitem{Banados:1992gq}
  M.~Banados, M.~Henneaux, C.~Teitelboim and J.~Zanelli,
  Phys.\ Rev.\ D {\bf 48} (1993) 1506
   Erratum: [Phys.\ Rev.\ D {\bf 88} (2013) 069902]
  doi:10.1103/PhysRevD.48.1506, 10.1103/PhysRevD.88.069902
  [gr-qc/9302012].

\bibitem{Anninos:2008qb}
  D.~Anninos,
  JHEP {\bf 0909} (2009) 075
  doi:10.1088/1126-6708/2009/09/075
  [arXiv:0809.2433 [hep-th]].

\bibitem{Clement:2009gq}
  G.~Clement,
  Class.\ Quant.\ Grav.\  {\bf 26} (2009) 105015
  doi:10.1088/0264-9381/26/10/105015
  [arXiv:0902.4634 [hep-th]].

\bibitem{Tonni:2010gb}
  E.~Tonni,
  JHEP {\bf 1008} (2010) 070
  doi:10.1007/JHEP08(2010)070
  [arXiv:1006.3489 [hep-th]].


\bibitem{Compere:2008cw}
  G.~Compere, S.~Detournay and M.~Romo,
  Phys.\ Rev.\ D {\bf 78} (2008) 104030
  doi:10.1103/PhysRevD.78.104030
  [arXiv:0808.1912 [hep-th]].

\bibitem{Detournay:2012dz}
  S.~Detournay and M.~Guica,
  JHEP {\bf 1308} (2013) 121
  doi:10.1007/JHEP08(2013)121
  [arXiv:1212.6792 [hep-th]].
   

\bibitem{Karndumri:2013dca}
  P.~Karndumri and E.~O.~Colgáin,
  JHEP {\bf 1310} (2013) 094
  doi:10.1007/JHEP10(2013)094
  [arXiv:1307.2086 [hep-th]].

\bibitem{Gurses:1994bjn}
  M.~Gurses,
  Class.\ Quant.\ Grav.\  {\bf 11} (1994) no.10,  2585.
  doi:10.1088/0264-9381/11/10/017

\bibitem{Barnich:2005kq}
  G.~Barnich and G.~Compere,
  Phys.\ Rev.\ Lett.\  {\bf 95} (2005) 031302
  doi:10.1103/PhysRevLett.95.031302
  [hep-th/0501102].


\bibitem{Jugeau:2010nq}
  F.~Jugeau, G.~Moutsopoulos and P.~Ritter,
  Class.\ Quant.\ Grav.\  {\bf 28} (2011) 035001
  doi:10.1088/0264-9381/28/3/035001
  [arXiv:1007.1961 [hep-th]].

\bibitem{Lloyd}
S. Lloyd,
Nature 406 (2000), no. 6799 1047-1054.

\bibitem{Alishahiha:2017hwg}
  M.~Alishahiha, A.~Faraji Astaneh, A.~Naseh and M.~H.~Vahidinia,
  JHEP {\bf 1705} (2017) 009
  doi:10.1007/JHEP05(2017)009
  [arXiv:1702.06796 [hep-th]].

\bibitem{Guo:2017rul}
  W.~D.~Guo, S.~W.~Wei, Y.~Y.~Li and Y.~X.~Liu,
  Eur.\ Phys.\ J.\ C {\bf 77} (2017) no.12,  904
  doi:10.1140/epjc/s10052-017-5466-5
  [arXiv:1703.10468 [gr-qc]].

\bibitem{Ghodrati:2017roz}
  M.~Ghodrati,
  Phys.\ Rev.\ D {\bf 96} (2017) no.10,  106020
  doi:10.1103/PhysRevD.96.106020
  [arXiv:1708.07981 [hep-th]].


\bibitem{Qaemmaqami:2017lzs}
  M.~M.~Qaemmaqami,
  Phys.\ Rev.\ D {\bf 97} (2018) no.2,  026006
  doi:10.1103/PhysRevD.97.026006
  [arXiv:1709.05894 [hep-th]].

  
  

\end{thebibliography}
\end{document}